\documentclass[aps,prd,nofootinbib,twocolumn,superscriptaddress,preprintnumbers,balancelastpage,longbibliography,10pt]{revtex4-2}
\usepackage{aas_macros}

\usepackage{amsmath}
\usepackage{orcidlink}
\usepackage{graphicx}
\usepackage{dcolumn}
\usepackage{bm}
\usepackage{xcolor}
\usepackage{booktabs}
\usepackage{siunitx}    
\usepackage{makecell}   

\usepackage[normalem]{ulem}
\usepackage{amssymb}

\renewcommand{\d}[1]{\textrm{d}#1} 

\newcommand{\sigmaobs}{\sigma_{\!\mathcal{L}}}
\newcommand{\sigmacat}{\sigma_{\rm cat}}
\newcommand{\sigmatrue}{\sigma}
\newcommand{\sigmav}{\langle\sigma v\rangle}

\begin{document}

\makeatletter
\typeout{Column width: \the\columnwidth}
\typeout{Text width: \the\textwidth}
\makeatother

\title{Systematic Uncertainties and Their Impact on Gamma-Ray Searches \\ for Dark Matter in Dwarf Galaxies}

\author{Toni Bertólez-Martínez\orcidlink{0000-0002-4586-6508}}
    \email{bertolezmart@wisc.edu}

\author{Dan Hooper\orcidlink{0000-0001-8837-4127}}
    \email{dwhooper@wisc.edu}

\author{Arifa Khatee Zathul \orcidlink{0000-0002-8735-8579}}
    \email{arifa@wisc.edu}
    
\affiliation{
    Department of Physics, Wisconsin IceCube Particle Astrophysics Center, University of Wisconsin, Madison, Wisconsin 53706, USA
}

\begin{abstract}

Gamma-ray observations of Milky Way dwarf galaxies provide some of the strongest constraints on dark matter annihilation and are approaching the sensitivity required to test dark matter interpretations of the Galactic Center Gamma-Ray Excess. The results, however, depend critically on the inferred dark matter distributions in these systems, conventionally described by their $J$-factors and spatial profiles. We review several sources of systematic uncertainty in dwarf galaxy $J$-factor determinations, including stellar membership, unresolved binaries, departures from equilibrium and spherical symmetry, and assumptions regarding the stellar and dark matter density profiles. Most $J$-factor determinations reported in the literature neglect several of these uncertainties, and the resulting measurements are often more dispersed than their quoted uncertainties would suggest. This indicates uncertainties of order $\Delta\log_{10}J\sim0.5$ or larger may be warranted for many of these systems. Using an ensemble of simulated Fermi-LAT observations of a stacked sample of 27 dwarf galaxies, we quantify how these unmodeled uncertainties affect the resulting constraints on the dark matter annihilation cross section. We find that adding an uncertainty of $\Delta\log_{10}J\sim 0.5$ or 1.0 in quadrature weakens the median upper limits of a stacked analysis by factors of 1.4 or 3.2, respectively. Furthermore, underestimating these uncertainties can cause such analyses to exclude the true annihilation cross section more frequently than the nominal confidence level would imply. Such undercoverage is especially pronounced in stacked analyses and persists even when the adopted and true $J$-factor uncertainties agree. Mismodeling spatially extended halos as point sources, as is sometimes done, further strengthens the resulting limits and increases the probability that true signals will be incorrectly excluded.

\end{abstract}

\maketitle


\section{Introduction}

If dark matter (DM) is a thermal relic of the early universe, the particles that constitute this substance are generally expected to possess approximately weak-scale masses and couplings~\cite{Bertone:2016nfn}. Such WIMPs -- weakly interacting massive particles -- have, however, been strongly constrained in recent years by the null results of direct~\cite{LZ:2024zvo, XENON:2025vwd, PandaX:2024qfu, CRESST:2019jnq, PICO:2019vsc} and indirect~\cite{Planck:2018vyg,John:2021ugy,Circiello:2026inp,DelaTorreLuque:2024ozf} searches. Nevertheless, many WIMP models remain consistent with existing constraints~\cite{Bertone:2018krk}.

The WIMP paradigm has received an intriguing measure of empirical support from the discovery and scrutiny of the Galactic Center Gamma-Ray Excess (GCE), as observed by the Fermi Large Area Telescope (LAT)~\cite{DiMauro:2026fnp,Cholis:2021rpp,DiMauro:2021raz} (for early work, see Refs.~\cite{Goodenough:2009gk,Hooper:2010mq,Hooper:2011ti,Abazajian:2012pn,Hooper:2013rwa,Gordon:2013vta,Daylan:2014rsa,Calore:2014xka,Fermi-LAT:2015sau}). The spectrum, angular distribution, and normalization of this signal are each consistent with expectations from annihilating WIMPs. For the case of annihilations to $b\bar{b}$, the excess is well fit by DM with a mass of $m_X \sim 35-65 \, {\rm GeV}$ and an annihilation cross section of $\langle \sigma v \rangle \sim (1-3) \times 10^{-26} \, {\rm cm}^3/{\rm s}$, distributed in the Inner Galaxy with $\rho \propto r^{-1.2}$. This cross section is consistent with that expected for DM in the form of a generic thermal relic, $\langle \sigma v \rangle \approx 2.2 \times 10^{-26} \, {\rm cm}^3/{\rm s}$, assuming velocity-independent annihilation~\cite{Steigman:2012nb}.

The leading astrophysical interpretation of the GCE is that it could be generated by a large population of faint and centrally located millisecond pulsars (MSPs)~\cite{Hooper:2010mq,Abazajian:2010zy,Hooper:2011ti,Abazajian:2012pn,Gordon:2013vta,Cholis:2014lta,Yuan:2014rca,Petrovic:2014xra,Brandt:2015ula,Malyshev:2024obk,Kuvatova:2024bdn}. Although pulsars are observed to produce gamma-ray spectra similar to that of the excess~\cite{Baltz:2006sv,Hooper:2010mq,Hooper:2011ti,Abazajian:2012pn,Gordon:2013vta}, several arguments disfavor this interpretation. 

First, very few MSPs have been detected in the inner kiloparsecs of the Milky Way. This implies that, if MSPs generate the excess, they must be systematically fainter than those observed in other environments, requiring a population of $\sim 10^5$ or more sources~\cite{Holst:2024fvb,Amerio:2024qor}. Recent studies of the Fermi data employing machine learning techniques likewise find no evidence that unresolved pulsars or other gamma-ray point sources are responsible for the excess~\cite{List:2025qbx} (see also Refs.~\cite{Leane:2019xiy,Leane:2020nmi,Leane:2020pfc,Zhong:2019ycb}).

Second, relatively few low-mass X-ray binaries (LMXBs) have been observed in the direction of the Inner Galaxy~\cite{Cholis:2014lta}. LMXBs are the primary progenitors of MSPs~\cite{Alpar:1982dec,Archibald:2009jun}, and their observed abundance suggests that MSPs account for only $\sim 4-11\%$ of the gamma-ray emission associated with the excess~\cite{Haggard:2017lyq}.

Third, neutron stars typically receive natal kicks of order $\sim 10^2~{\rm km/s}$ at the time of their formation. These kicks should broaden the MSP distribution relative to the stellar population, making it difficult to reproduce the observed central concentration of the gamma-ray emission, particularly within the innermost degree around the Galactic Center~\cite{Boodram:2022lpn}. More broadly, the excess extends to at least $20^{\circ}$ north and south of the Galactic Plane, well beyond the extent of the stellar bulge~\cite{DiMauro:2026fnp}. Even after accounting for natal kicks, this extent is difficult to reconcile with pulsars or other stellar populations (see also Refs.~\cite{Zhong:2024vyi,McDermott:2022zmq,Cholis:2021rpp,DiMauro:2021raz, Song:2024iup}).

Despite these arguments, an MSP origin of the GCE cannot be entirely excluded. Such an explanation, however, would require an exotic population with a systematically faint luminosity function, relatively few associated LMXBs, and a spatial distribution following $n\propto r^{-2.4}$ out to at least several kpc from the Galactic Center. It is therefore essential to pursue other search strategies that could confirm or rule out DM interpretations of the GCE. The most important of these are gamma-ray observations of the Milky Way's dwarf galaxy population.

If the GCE is generated by annihilating DM, each dwarf galaxy should produce a gamma-ray signal with the same spectral shape (see, however, Ref.~\cite{Berlin:2025fwx}). Although dwarf spheroidal galaxies are expected to yield a much smaller gamma-ray flux from DM than the Inner Galaxy (by a factor of $\sim 10^3-10^4$ or more), these systems contain little gas and have undergone essentially no recent star formation, and thus generate much lower intrinsic astrophysical backgrounds. Thus, although a signal from annihilating DM would almost certainly be detected first from the Inner Galaxy, subsequent observations of dwarf galaxies could very well be the most convincing way to confirm or exclude DM interpretations of that signal.

Fermi-LAT searches for DM annihilation in dwarf galaxies are now approaching the sensitivity required to test DM interpretations of the GCE~\cite{Circiello:2026inp,DiMauro:2022hue}. Interpreting these results, however, can involve considerable challenges. In this article, we review some of the most important systematic uncertainties associated with the kinematic inference of dwarf galaxy $J$-factors and quantify their impact on indirect searches with Fermi. In Section II, we review the leading sources of uncertainty in $J$-factor inference and their treatment in the literature. In Section III, we describe our methods. Sections IV and V present results for simulations of a single dwarf galaxy and for a stacked analysis of 27 dwarfs, respectively. We summarize our conclusions in Section VI.

\section{Dwarf Galaxy $J$-factors and their Uncertainties}\label{sec:systematics}

The gamma-ray flux from DM annihilation, integrated over a solid angle, $\Delta \Omega$, is given by
\begin{align}\label{eq:DM-flux}
\frac{\d N_{\gamma}}{\d E_{\gamma}}(E_{\gamma}, \Delta \Omega) &= \frac{\langle \sigma v \rangle}{8 \pi m_X^2} \, \frac{\d N_{\gamma}}{\d E_{\gamma}}\bigg|_{\rm ann}  \int_{\!\Delta \Omega} \int_{\rm los} \rho_X^2 \, \d l \, \d\Omega  \nonumber \\
&= \frac{\langle \sigma v \rangle}{8 \pi m_X^2}  \,  \frac{\d N_{\gamma}}{\d E_{\gamma}}\bigg|_{\rm ann} \,J(\Delta \Omega), 
\end{align}
where $\langle \sigma v \rangle$ is the velocity-averaged DM annihilation cross section, $m_X$ is the mass of the DM particle, and $\d N_{\gamma}/\d E_{\gamma}|_{\rm ann}$ is the spectrum of gamma rays produced per annihilation. The $J$-factor, $J(\Delta \Omega)$, is defined as the square of the DM density, $\rho_X$, integrated over a solid angle, $\Delta \Omega$, and along the line of sight, $l$. 

The $J$-factor of a given dwarf galaxy can be inferred from its resolved stellar kinematics. Such analyses combine the measured positions and line-of-sight velocities of likely member stars (together with metallicities and proper motions, when available) with models of the stellar surface-density profile, DM halo, and stellar velocity-anisotropy profile. Conventional analyses assume that the dwarf is a spherical, pressure-supported system in dynamical equilibrium. The spherical Jeans equation is then used to determine the stellar velocity-dispersion profile, which is projected along the line of sight and compared with the observed stellar velocities. An unbinned likelihood for the individual stellar measurements is then explored using Bayesian sampling or, less commonly, a frequentist profile-likelihood construction. For each allowed set of halo parameters, the density-squared integral is evaluated, producing a posterior for $J$. The result is usually quoted as a median value and standard deviation in $\log_{10}J$ within a specified angular aperture, commonly $0.5^\circ$~\cite{Geringer-Sameth:2014qqa,Bonnivard:2015xpq,Chiappo:2016xfs} (for a review, see Ref.~\cite{Strigari:2018utn}).

The precision of $J$-factor determinations is limited in part by the size and quality of the available stellar-kinematic samples. Classical dwarfs often have hundreds or thousands of measured member stars, making it possible to obtain relatively well-constrained $J$-factors for these systems. In contrast, estimates for ultra-faint systems can sometimes be based on only a few tens of stars, yielding results with large uncertainties that depend strongly on the membership assignments of individual stars and on the adopted priors. 

In addition to these statistical limitations, gamma-ray searches for DM annihilation in dwarf galaxies can be biased or otherwise inaccurate as a result of several systematic effects, some of which include the following:

\begin{enumerate}
    \item{{\bf Neglecting Spatial Extension:} In many gamma-ray analyses, dwarf galaxies are treated as point sources, neglecting their spatial extent. In reality, many dwarfs -- especially the nearest systems, which tend to have the largest $J$-factors -- have apparent sizes comparable to or larger than the point spread function (PSF) of Fermi-LAT~\cite{DiMauro:2022hue,Raman:2026zbr}. By treating dwarf galaxies as point sources, such analyses risk assigning part of their gamma-ray emission to astrophysical backgrounds, yielding upper limits that are artificially stringent by factors of $\sim 1.5-1.8$ in stacked analyses~\cite{DiMauro:2022hue}.}
    \item{{\bf Mismodeling of Halo Profiles:} The DM distribution in a given dwarf galaxy is typically assumed to follow a specified parametric form, such as a tidally truncated Navarro-Frenk-White (NFW) profile. In reality, the DM distribution can vary from dwarf to dwarf and could, in some cases, feature dense cusps or approximately constant-density cores~\cite{Read:2015sta,Klop:2016lug,Genina:2018abc}. The presence of such cores could weaken DM annihilation constraints by factors of a few~\cite{Klop:2016lug}.}
    \item{{\bf Mismodeling of Stellar Profiles:}
    To convert the projected stellar density into a three-dimensional stellar distribution, one generally fits a specified functional form to the observed stellar surface-density profile. The standard choice for this is a Plummer profile~\cite{Plummer:1911mar,Pace:2018tin}, which is not always sufficiently flexible to describe the data adequately and can consequently lead to underestimated $J$-factor uncertainties~\cite{Bonnivard:2014kza,Chiappo:2018mlt}.   
    On another note, recent work has pointed out that the choice of stellar density profile can introduce systematic uncertainties of up to an order of magnitude in a dwarf's half-light radius and dynamical mass~\cite{Walker:2026abc,Splawska:2026kti}.}
    \item{{\bf Departures from Sphericity:} The DM halos of dwarf galaxies are typically assumed to be spherical, while simulations indicate that many are significantly triaxial. This affects not only the angular distribution of the predicted annihilation signal but also the overall normalization of the inferred $J$-factor~\cite{Bonnivard:2014kza,Genina:2018abc,Klop:2016lug,Hayashi:2016kcy}. Modeling an axisymmetric dwarf with a spherical NFW profile can bias the inferred value of $\log_{10} J$ by up to $\sim 0.4$. This effect can dominate the $J$-factor uncertainties among the classical dwarf sample~\cite{Bonnivard:2015xpq}.}
    \item{{\bf Departures from Equilibrium:} The Jeans analyses employed in typical $J$-factor determinations implicitly assume that the dwarf is in dynamical equilibrium. In reality, dwarfs interact with the gravitational potential of the Milky Way and tidal effects can modify their stellar dynamics and drive them away from equilibrium. For example, Refs.~\cite{Bonnivard:2015xpq,Walker:2026abc} suggest that the morphologies of Hercules and Ursa Major II are consistent with ongoing tidal disruption. Simulations indicate that departures from equilibrium can introduce an additional uncertainty as large as $\Delta \log_{10} J \sim 0.1$ for some dwarf galaxies~\cite{Tchiorniy:2026abc}.}
    \item{{\bf The Presence of Binary Populations:} In ultra-faint dwarf galaxies, unresolved binary stellar systems can increase the measured stellar velocity dispersion above that generated by the gravitational potential of the DM halo. For instance, in the case of Reticulum II, this effect may have inflated the inferred value of $\log_{10} J$ by $\sim 0.3$~\cite{Minor:2019abc}. In extreme cases, objects classified as ultra-faint dwarf galaxies could actually be diffuse globular clusters with little actual DM content.}
    \item{{\bf Foreground Contamination:} Stars along the line of sight to a given dwarf can sometimes be mistakenly identified as members of that system, biasing the Jeans analysis. Inappropriate membership cuts or sample-selection criteria can cause $J$-factors to be overestimated, especially for dwarfs with relatively few measured member stars~\cite{Bonnivard:2016abc,Ichikawa:2017rph}.}
    \item{{\bf Non-Gaussian $J$-factor Likelihoods:} Stacked dwarf analyses typically treat the uncertainties in the $J$-factors as log-normal. This approximation is not expected to capture the full likelihood profile associated with these determinations, which can feature significantly heavier tails~\cite{Pace:2018tin, Alvarez:2020cmw,Horigome:2022gge}. 
    For this reason, Ref.~\cite{Pace:2018tin} points out that caution is needed with the nominal $J$-factors of Draco II, Grus I, Leo IV, Leo V, Pegasus III, and Pisces II.}
\end{enumerate}

\begin{figure*}
    \centering
    \includegraphics[width=1\linewidth]{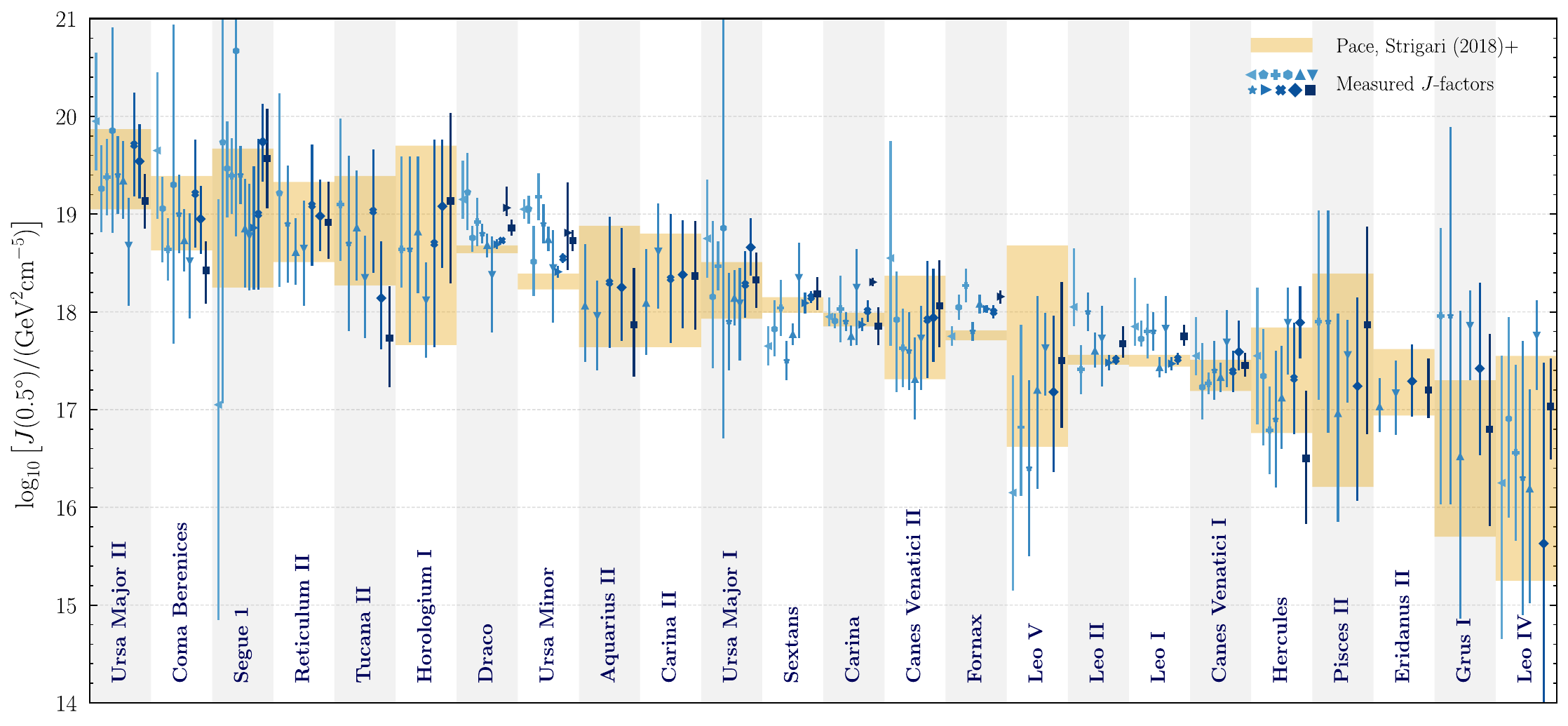}
    \caption{Measurements of dwarf galaxy $J$-factors reported in the literature. For each galaxy, the yellow band shows the 1$\sigma$ interval for the $J$-factor integrated within a circle of radius $0.5^{\circ}$, as reported in Refs.~\cite{Pace:2018tin,Boddy:2019qak}. We adopt these $J$-factors and their uncertainties as the default values in our analysis. Here, we show only those dwarf galaxies with three or more different reported measurements. Each set of measurements is sorted in chronological order, with darker colors denoting more recent measurements. The papers corresponding to each of the colors and symbols are 
    \includegraphics[height=1.5ex]{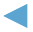}~\cite{Bonnivard:2015xpq}, 
    \includegraphics[height=1.5ex]{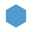}~\cite{Chiappo:2016xfs}, 
    \includegraphics[height=1.5ex]{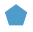}~\cite{Bonnivard:2016abc},
    \includegraphics[height=1.5ex]{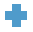}~\cite{Evans:2016xwx},
    \includegraphics[height=1.5ex]{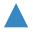}~\cite{Pace:2018tin},
    \includegraphics[height=1.5ex]{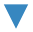}~\cite{Fattahi:2018mrnas},
    \includegraphics[height=1.5ex]{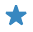}~\cite{Calore:2018sbp},
    \includegraphics[height=1.5ex]{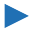}~\cite{Alvarez:2020cmw}, 
    \includegraphics[height=1.5ex]{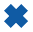}~\cite{DiMauro:2022hue},
    \includegraphics[height=1.5ex]{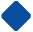}~\cite{Horigome:2022gge}, and
    \includegraphics[height=1.5ex]{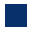}~\cite{Raman:2026zbr}. When an article reports more than one $J$-factor, the different measurements are combined for visualization by averaging their central values, with error bars that incorporate the dispersion among them in quadrature. \textit{Notice that, for some of these dwarfs, the reported $J$-factor measurements are significantly more dispersed than would be expected from the sizes of their individual error bars. We further emphasize that these determinations do not account for many of the systematic uncertainties considered in this study and therefore underestimate the uncertainties on the associated $J$-factors.}
    }
    \label{fig:Jcomparison}
\end{figure*}

In Fig.~\ref{fig:Jcomparison}, we show the measured $J$-factors and estimated uncertainties for a sample of dwarf galaxies, as reported in previous studies. The darker blue (and further-right) points correspond to more recent results. Notice that, for many dwarf galaxies, the dispersion among the reported measurements is substantially larger than would be expected from their quoted uncertainties. In many cases, this dispersion cannot be accounted for by improvements in data quality or analysis techniques, but is instead indicative of systematic uncertainties that are not fully captured by the reported error bars. We emphasize that each of these determinations relies on at least some of the assumptions and approximations discussed above and is thus likely to underestimate the true uncertainty in the corresponding $J$-factors.

\section{Methods and sample selection}

A typical gamma-ray analysis searching for DM annihilation in a collection of dwarf galaxies begins by selecting Fermi-LAT photons from a region of interest (ROI) around each dwarf galaxy and constructing a Poisson likelihood binned in spatial position and energy. The model for each ROI includes the Galactic and isotropic diffuse backgrounds, cataloged gamma-ray sources, and a putative DM component centered on the dwarf. The normalizations, and in some cases the spectral parameters, of the most relevant background components are fitted as nuisance parameters~\cite{Fermi-LAT:2011vow,Fermi-LAT:2015att}.

\begin{table*}[htbp]
\setlength{\tabcolsep}{5pt}
\centering
\begin{tabular}{lccccc}
\toprule
\textbf{Name} & \textbf{Distance (kpc)} & \textbf{$\log_{10}J_{\text{tot}}$} & \textbf{$\log_{10}J(0.5^\circ)$} & \textbf{$\theta_{68}$ (deg)} & \textbf{Shape parameters} \\
\midrule
\multicolumn{5}{l}{\textbf{cored-NFW classical dwarf galaxies}} & \multicolumn{1}{c}{$\log_{10}\rho_s,\, r_s,\ \ r_t,\ \ n,\ \ r_c,\ \ \delta\ \ \ $} \\
\midrule
Carina & 105.0 & $18.01 \pm 0.11$ & $17.92 \pm 0.07$ & 0.36 & 7.16, 1.66, 1.61, 0.53, 0.60, 4.21 \\
Draco & 76.0 & $18.73 \pm 0.03$ & $18.64 \pm 0.04$ & 0.35 & 7.34, 1.68, 1.11, 0.54, 0.19, 4.16 \\
Fornax & 147.0 & $18.0 \pm 0.07$ & $17.76 \pm 0.05$ & 0.59 & 7.07, 2.75, 2.27, 0.86, 1.94, 4.39 \\
Leo I & 254.0 & $17.52 \pm 0.06$ & $17.5 \pm 0.06$ & 0.12 & 7.36, 1.48, 1.37, 0.52, 0.41, 4.26 \\
Leo II & 233.0 & $17.51 \pm 0.05$ & $17.51 \pm 0.05$ & 0.07 & 7.61, 0.97, 0.54, 0.51, 0.23, 4.21 \\
Sextans & 86.0 & $18.15 \pm 0.06$ & $18.07 \pm 0.08$ & 0.35 & 7.41, 1.13, 1.26, 0.59, 0.53, 4.22 \\
Ursa Minor & 76.0 & $18.55 \pm 0.05$ & $18.31 \pm 0.08$ & 0.59 & 7.30, 2.17, 1.11, 0.79, 1.40, 4.29 \\
\midrule
\multicolumn{5}{l}{\textbf{NFW ultra-faint dwarf galaxies}} & 
\multicolumn{1}{c}{$\log_{10}\rho_s,\,   r_s,\, \ \ \  r_t\quad $} \\
\midrule
Aquarius II & 108.0 & $18.30 \pm 0.67$ & $18.26 \pm 0.62$ & 0.19 & 7.55, 1.01, 10.41 \\
Canes Venatici I & 218.0 & $17.39 \pm 0.21$ & $17.35 \pm 0.16$ & 0.17 & 7.02, 1.75, 21.56 \\
Canes Venatici II & 160.0 & $17.92 \pm 0.60$ & $17.84 \pm 0.53$ & 0.25 & 7.07, 2.02, 19.61 \\
Carina II & 36.0 & $18.34 \pm 0.66$ & $18.22 \pm 0.58$ & 0.38 & 7.32, 0.73, 19.61 \\
Coma Berenices & 44.0 & $19.21 \pm 0.55$ & $19.01 \pm 0.38$ & 0.58 & 7.46, 1.24, 4.82 \\
Draco II & 22.0 & $18.93 \pm 1.55$ & $18.93 \pm 1.54$ & 0.19 & 4.90, 0.30, 1.00 \\
Eridanus II & 380.0 & $17.28 \pm 0.34$ & $17.28 \pm 0.34$ & 0.081 & 7.42, 1.31, 25.00 \\
Grus I & 120.0 & $16.50 \pm 0.80$ & $16.50 \pm 0.80$ & 0.12 & 4.50, 0.60, 1.00 \\
Hercules & 132.0 & $17.32 \pm 0.57$ & $17.30 \pm 0.54$ & 0.11 & 7.39, 0.69, 7.34 \\
Horologium I & 79.0 & $18.70 \pm 1.06$ & $18.68 \pm 1.02$ & 0.13 & 8.03, 0.56, 7.46 \\
Hydrus I & 28.0 & $18.93 \pm 0.57$ & $18.65 \pm 0.32$ & 0.60 & 7.35, 0.92, 1.98 \\
Leo IV & 154.0 & $16.40 \pm 1.15$ & $16.40 \pm 1.15$ & 0.079 & 4.40, 0.84, 1.00 \\
Leo V & 178.0 & $17.65 \pm 1.03$ & $17.65 \pm 1.03$ & 0.016 & 6.89, 0.45, 1.00 \\
Pisces II & 182.0 & $17.30 \pm 1.09$ & $17.30 \pm 1.09$ & 0.032 & 6.93, 0.57, 1.00 \\
Reticulum II & 30.0 & $19.09 \pm 0.62$ & $18.92 \pm 0.41$ & 0.51 & 7.54, 0.83, 2.62 \\
Sagittarius II & 69.0 & $17.48 \pm 1.23$ & $17.35 \pm 1.36$ & 0.092 & 7.39, 0.59, 3.12 \\
Segue 1 & 23.0 & $19.00 \pm 0.77$ & $18.96 \pm 0.71$ & 0.17 & 8.30, 0.19, 1.14 \\
Tucana II & 58.0 & $19.03 \pm 0.63$ & $18.83 \pm 0.56$ & 0.57 & 7.31, 1.65, 6.74 \\
Ursa Major I & 97.0 & $18.28 \pm 0.34$ & $18.22 \pm 0.29$ & 0.24 & 7.42, 1.11, 9.91 \\
Ursa Major II & 32.0 & $19.71 \pm 0.53$ & $19.46 \pm 0.41$ & 0.74 & 7.61, 1.25, 5.29 \\
\bottomrule
\end{tabular}
\caption{The parameters of the dwarf spheroidal galaxies included in our simulations, as reported in Refs.~\cite{Pace:2018tin,Boddy:2019qak,DiMauro:2022hue}. The $J$-factors are given in units of $\rm GeV^2/cm^5$, $\rho_s$ in $M_\odot/{\rm kpc}^3$, and $r_s$, $r_t$, and $r_c$ in kpc. For those ultra-faint dwarfs with $r_t=1.00$~kpc, the inferred tidal radius is sufficiently large relative to the spatial extent of the dwarf that the fit does not meaningfully constrain $r_t$; its precise value therefore has a negligible impact on our analysis.}
\label{tab:dsphs}
\end{table*}

The information from different dwarfs is then combined at the likelihood level, rather than by simply summing their photon counts. For a specified DM mass and annihilation final state, the annihilation cross section is treated as a common parameter across all dwarfs, while each target retains its own background parameters and $J$-factor. The uncertainty in each $J$-factor is incorporated through an additional likelihood term -- often approximated as Gaussian in $\log_{10}J$ -- and is profiled or marginalized over when constructing the joint likelihood. Dwarfs with large $J$-factors, well-measured stellar kinematics, and relatively low gamma-ray backgrounds therefore contribute the greatest statistical weight. The combined likelihood is scanned as a function of the DM annihilation cross section. In the absence of a significant collective excess, the resulting likelihood profile is used to derive a 95\% confidence-level upper limit on $\langle \sigma v \rangle$ as a function of the DM mass and annihilation channel. This methodology was used, for example, in the canonical six-year Pass 8 analysis~\cite{Fermi-LAT:2015att} and, with updated data, target lists, and $J$-factors, in the more recent analysis based on more than 15 years of Fermi-LAT observations~\cite{Circiello:2026inp} (see also Refs.~\cite{McDaniel:2023bju,DiMauro:2022hue,Fermi-LAT:2011vow}).

 In this study, we simulate the Fermi-LAT data and analyze that simulated dataset using the \texttt{FermiTools} (v2.5.1)~\cite{fermitools} and the \texttt{fermipy} library (v1.4.0)~\cite{wood2017fermipyopensourcepythonpackage}. In most respects, we closely follow the procedures adopted in previous Fermi analyses~\cite{Circiello:2026inp,McDaniel:2023bju,Fermi-LAT:2013sme,Fermi-LAT:2015att,Fermi-LAT:2015ycq,Fermi-LAT:2016uux}. In particular, our baseline model includes events with energies between 500~MeV and 1~TeV, selected using the \texttt{P8R3\_SOURCE\_V3} instrument response functions and divided into four PSF event classes, each with its own corresponding isotropic diffuse template. The exposure is calculated for 17.4 years, spanning August 4, 2008 to January 14, 2026. Each ROI is taken to be $10^\circ\times10^\circ$, divided into $0.08^\circ$ bins and 8 energy bins per decade. The baseline model includes the Galactic diffuse emission template, \texttt{gll\_iem\_v07.fits}, and corresponding isotropic diffuse emission. For simplicity, we do not include 4FGL catalog sources in our simulated datasets.

In Table~\ref{tab:dsphs}, we list several of the relevant quantities for each of the dwarf galaxies included in our analysis. These dwarfs were selected to match the ``Measured'' sample adopted in Ref.~\cite{Circiello:2026inp}. Each has a $J$-factor that was determined through a Jeans analysis and shows no evidence of significant contamination by background gamma-ray sources. We further restrict this list to those dwarfs for which shape information is provided in Ref.~\cite{Pace:2018tin}. For each dwarf, we consider both the total $J$-factor and the $J$-factor integrated within $0.5^\circ$, adopting values from Ref.~\cite{DiMauro:2022hue} when available, and otherwise from Refs.~\cite{Pace:2018tin,Boddy:2019qak}. Table~\ref{tab:dsphs} also illustrates the angular extent of each dwarf by reporting the value of $\theta_{68}$, defined such that $J(\theta_{68})=0.68\,J_{\rm tot}$. This quantity is included only for illustrative purposes; in our simulations, we adopt realistic DM profiles following Ref.~\cite{DiMauro:2022hue}. 

For the classical dwarfs, we adopt cored and tidally truncated NFW profiles, parameterized as follows~\cite{Read:2015sta,Read:2018pft}:
\begin{equation}
    \rho_{\mathrm{cNFWt}}(r)=\left\{\begin{array}{l l}\rho_{\mathrm{cNFW}}(r)&r\leqslant r_{t}\\ \rho_{\mathrm{cNFW}}(r_{t})(r/r_{t})^{-\delta}&r>r_{t},\end{array}\right.
\end{equation}
where
\begin{align}
     \rho_{\mathrm{cNFW}}(r) = f^n \rho_{\mathrm{NFW}} + \frac{n f^{n-1}(1 - f^2)}{4\pi r^2 r_c} M_{\mathrm{NFW}},  
\end{align}
with $f = \tanh(r/r_c)$, and $r_c$ denoting the core radius. Here, $\rho_{\rm NFW}$ and $M_{\rm NFW}$ are, respectively, the Navarro-Frenk-White density and enclosed-mass profiles~\cite{Navarro:1996gj}, 
\begin{align}\label{eq:NFW-profile}
 \rho_{\mathrm{NFW}}(r)&=\rho_{s}\frac{r_{s}}{r}\frac{1}{(1+r/r_{s})^{2}}  \nonumber \\
M_{\rm NFW}(r) &= 4 \pi \int_0^r \d r' (r')^2 \rho_{\rm NFW}(r').
\end{align}

For the ultra-faint dwarfs, we instead adopt a tidally truncated NFW profile, described by Eq.~\ref{eq:NFW-profile} for $r< r_t$ and set to $\rho=0$ for $r>r_t$.

We parameterize the asphericity of each dwarf in terms of the ratios of its three principal axes, $(a,b,c)$. An arbitrary ellipsoidal density profile can be obtained from the corresponding spherical profile by replacing the radius $r$ with the ellipsoidal radius:
\begin{equation}
    r = \sqrt{\left(\frac{x}{a}\right)^2+
              \left(\frac{y}{b}\right)^2+
              \left(\frac{z}{c}\right)^2}, 
\end{equation}
where $a\ge b\ge c$ and $abc=1$~\cite{DiMauro:2022hue}. Specifying the apparent shape of the dwarf as viewed from Earth also requires three orientation angles. For simplicity, we set these angles to zero, corresponding to an edge-on orientation in which the intermediate axis is aligned along our line of sight.

For each dwarf galaxy, we simulate the background within the ROI by drawing Poisson counts from the baseline model. We then inject a simulated DM signal on top of this background. The injected DM source is modeled using the gamma-ray spectra provided in Ref.~\cite{Cirelli:2010xx}, normalized to $J_{\rm tot}$, and with a spatial shape that follows from the corresponding DM profile. Following injection, the source and background components are fitted, and a spectral energy distribution (SED) is determined for the dwarf galaxy, leaving the background normalization free to vary. In general, the SED analysis returns a Poisson likelihood in each energy bin, 
\begin{align}
\mathcal{L}_{\rm SED}(\mu, \theta | \mathcal{D}) = \prod_k \frac{\lambda_k^{n_k} e^{-\lambda_k}}{n_k!},
\end{align}
where $n_k(\mathcal{D})$ is the number of events observed in the $k$th bin of the dataset $\mathcal{D}$, and $\lambda_{k}(\mu, \theta)$ is the corresponding mean number of events predicted by the model, characterized by the signal parameters $\mu$ and nuisance parameters $\theta$.

As defined in Eq.~\ref{eq:DM-flux}, the DM signal is specified by three quantities: the annihilation cross section, mass, and annihilation channel. In this work, we focus on the case of $m_X=42\,\rm GeV$ and annihilations to $b\bar{b}$, as motivated by the observed spectrum of the GCE~\cite{DiMauro:2021qcf}. 

Among the nuisance parameters, the $J$-factor controls the normalization of the gamma-ray flux from DM annihilation. In this regard, we distinguish between two quantities. The observed value, $J_{\rm obs}$, is the central value as determined by the Jeans analysis and around which the $J$-factor likelihood is constructed. The true value, $J_{\rm true}$, is the physical $J$-factor used to normalize the injected gamma-ray signal. In our simulations, we set $J_{\rm true}$ to the central value reported in Table~\ref{tab:dsphs}.

For a single dwarf galaxy, the constraints on the $J$-factor are implemented in our likelihood through the term
\begin{align}
\label{eq:LJ}
\mathcal{L}_J(J|J_{\rm obs},\sigma_{\mathcal L}) = \frac{e^{  -(\log_{10} J-\log_{10} J_{\rm obs})^2/2\sigma_{\mathcal L}^2     }}{\ln(10) J_{\rm obs} \sqrt{2\pi} \sigma_{\mathcal L}}, 
\end{align}
where $J$ is the $J$-factor profiled over in the gamma-ray analysis and
$\sigma_{\mathcal L}$ is the uncertainty assigned to the $J$-factor
inference. In our baseline analysis, we adopt the values of $J_{\rm obs}$ and
$\sigma_{\mathcal L}$ from Table~\ref{tab:dsphs}.

The total likelihood for a single dwarf is given by
\begin{align}\label{eq:Lone}
    \mathcal{L}_i(\sigmav,J, \alpha\,&|\,\mathcal{D},J_{\rm obs},\sigma_{\mathcal L}) = \\
    & \ \mathcal{L}_J(J \, | \, J_{\rm obs},\sigma_{\mathcal L})\times \mathcal{L}_{\rm SED}(\sigmav, J,\alpha \,|\, \mathcal{D}),    \nonumber
\end{align}
where $\alpha$ denotes the nuisance parameters other than the $J$-factor normalization, such as spatial shape parameters. This can be extended to multiple dwarf galaxies by taking the product of the individual likelihoods,
\begin{align}\label{eq:Ljoint}
\mathcal{L}(\sigmav, \boldsymbol{J},\boldsymbol{\alpha}\,|\,&\mathcal{D},\boldsymbol{J_{\rm obs}},\boldsymbol{\sigma}_{\mathcal L}) =\\ & \prod_i \mathcal{L}_i(\sigmav, J_i,\alpha_i\,|\,\mathcal{D}_i,J_{{\rm obs},i},\sigma_{{\mathcal L},i}),
\nonumber
\end{align}
where the product is over the galaxies included in the stack, $\mathcal{L}_i$ is the likelihood of the $i$th galaxy (as calculated using Eq.~\ref{eq:Lone}), and bold symbols denote sets of parameters, e.g.,  $\boldsymbol{J} = \{J_i\}$.

Finally, given a likelihood $\mathcal{L}$, we define the test statistic (TS) for the significance of a DM signal as
\begin{equation}
    \mathrm{TS}= \max_{\sigmav}\left\{ 2\log\left(\frac{\mathcal{L}(\sigmav) }{\mathcal{L}(\sigmav=0) }\right)\right\}\, ,
\end{equation}
where nuisance parameters are profiled over. Assuming the applicability of Wilks' theorem (which we discuss later), the 95\% upper limit on $\sigmav$ is obtained for $\Delta\rm TS=2.71$. In this context, $\sigmav$ is the value of the cross section that yields the largest value of the test statistic.

\section{The Impact of Unmodeled Uncertainties on the Analysis of Individual Dwarf Galaxies}

Unaccounted-for systematics can increase the uncertainty in the determination of dwarf galaxy $J$-factors beyond those reported in the literature. By adopting these underestimated uncertainties, gamma-ray analyses of dwarf galaxies can be expected to produce constraints on DM annihilation that are artificially stringent. In this Section, we quantify the impact of these unmodeled uncertainties by focusing on the analysis of an individual dwarf galaxy, using Ursa Minor as an illustrative example. Ursa Minor is a ``classical dwarf'' with thousands of spectroscopically identified member stars. This makes its $J$-factor determination relatively robust against stellar membership contamination and the effects of unresolved binary populations~\cite{Bonnivard:2015xpq}. Its proximity, $D = 76\pm 4\, \rm kpc$~\cite{Pace:2018tin}, also contributes to its large $J$-factor, $\log_{10}\left(J_{\rm tot}/{\rm [GeV^2 cm^{-5}]}\right)=18.55\pm 0.05$, making Ursa Minor one of the most promising dwarf galaxies for DM annihilation searches.

Our knowledge of Ursa Minor and its treatment in DM annihilation analyses, however, leaves several important questions unanswered. Since this dwarf galaxy is extended and relatively nearby, its angular extent ($\theta_{68}= 0.59^\circ$ in our model) is larger than the Fermi-LAT PSF. It also has an elongated shape, with a projected ellipticity of $e\sim 0.56$~\cite{Bonnivard:2015xpq}, but it is not yet clear whether this system is flattened and oblate or triaxial~\cite{An:2022apr}, a distinction that could affect constraints on $\sigmav$ by up to 40\%~\cite{DiMauro:2022hue}. 

Recent studies have pointed out that determinations of Ursa Minor’s $J$-factor are subject to selection effects, with the inferred results depending strongly on the stellar populations included in the Jeans analysis~\cite{Yang:2025nov}. Recent measurements of the DM profile also exhibit a large dispersion in the inferred inner slope, depending on the spectroscopic catalog used, with the inferred central density varying by approximately an order of magnitude~\cite{Pascale:2026jun}. Furthermore, tailored simulations show that tidal forces from the Milky Way can cause Jeans analyses to overestimate the central density and underestimate the associated uncertainties~\cite{Tchiorniy:2026abc}. Tidal effects could also explain an excess of stars in the outskirts of the galaxy~\cite{Sestito:2023oct}, although other studies argue that this feature instead arises from the presence of two distinct stellar populations in Ursa Minor~\cite{Boyea:2026apr}. These considerations demonstrate that, even for a relatively well-measured classical dwarf galaxy such as Ursa Minor, there are likely substantial unmodeled systematic uncertainties. As an indication of this, notice that recent determinations of Ursa Minor’s $J$-factor have varied by significantly more than would be expected from their quoted uncertainties (see Fig.~\ref{fig:Jcomparison}).

\begin{figure}
    \centering \includegraphics[width=1\linewidth]{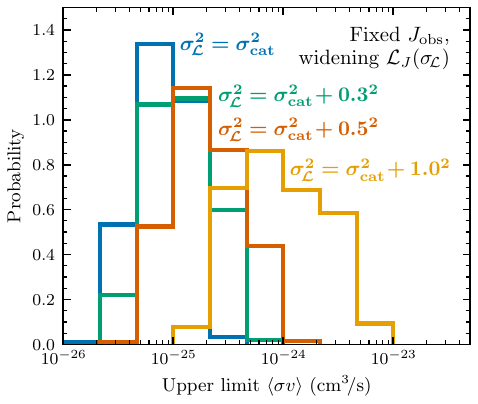}
    \caption{\textbf{Ursa Minor sensitivity.} The probability distribution of the 95\% confidence level upper limits on the DM annihilation cross section obtained from our simulated Fermi-LAT data from the ROI around Ursa Minor, assuming no signal from annihilating DM. \textit{As we increase the unmodeled uncertainty in this dwarf galaxy’s $J$-factor ($\sigmaobs > \sigma_{\rm cat}$, where $\sigma_{\rm cat}$ is the uncertainty quoted in Table~\ref{tab:dsphs}), the upper limits on $\sigmav$ become weaker and their distribution broadens.}}
    \label{fig:UMi-null-hyp}
\end{figure}

\begin{figure*}
    \centering
    \includegraphics[width=1\linewidth]{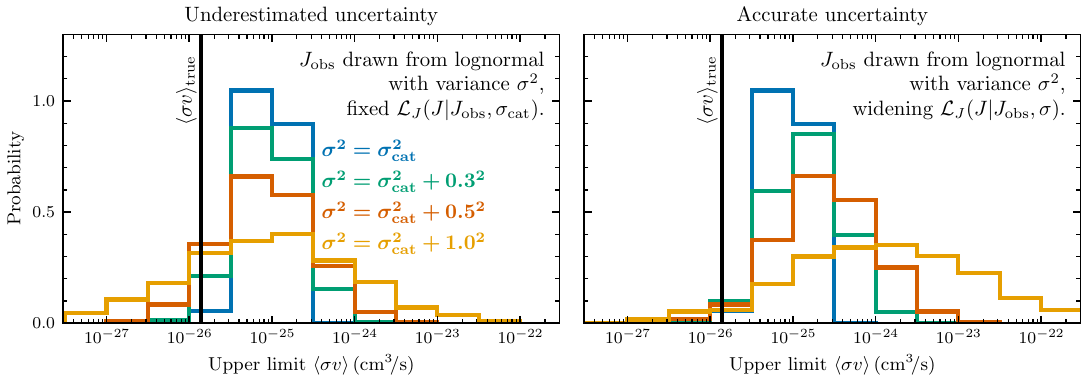}
    \caption{\textbf{GCE-like signal from Ursa Minor with increased uncertainty.} 
    Distributions of the 95\% confidence level upper limits on the DM annihilation cross section obtained from our simulations of Ursa Minor, including a DM signal with $\sigmav = 1.4\times 10^{-26} \, {\rm cm}^3/{\rm s}$. In the left panel, the observed $J$-factor, $J_{\rm obs}$, is drawn from a log-normal distribution with variance $\sigmatrue^2$, while the smaller uncertainty, $\sigmaobs^2=\sigmacat^2$, is adopted when calculating the likelihood. In the right panel, by contrast, we increase $\sigmaobs^2$ to match $\sigmatrue^2$. The blue histograms correspond to the standard baseline, while the green, red and yellow histograms correspond to progressively larger values of $\sigmatrue$. \textit{Underestimating the uncertainties associated with $J$-factor determinations can lead to artificially stringent upper limits and may even exclude signals that are present in the data. This effect can be mitigated by adopting larger values of $\sigma_{\mathcal L}$ that reflect the true uncertainty associated with the dwarf galaxy’s $J$-factor.}}
    \label{fig:UMi-true-signal}
\end{figure*}

In the following, we take an agnostic stance regarding the systematic uncertainties in Ursa Minor’s $J$-factor and simply increase the uncertainty associated with this determination. To account for these unmodeled uncertainties, we modify the width of the likelihood in Eq.~\ref{eq:LJ} as follows:
\begin{equation}\label{eq:LJvar}
    \mathcal{L}_J(J|J_{\rm obs},\sigmacat)\  \to \ 
    \mathcal{L}_J(J|J_{\rm obs},\sigmaobs), \,\,\,\,\,\,\, \sigmaobs\ge \sigmacat
\end{equation}
where $\sigma_{\rm cat}$ is the uncertainty quoted in Table~\ref{tab:dsphs}. In other words, choosing $\sigmaobs>\sigmacat$ makes an analysis more conservative by allowing for the possibility of unaccounted-for systematic uncertainties, such as those described in Section~\ref{sec:systematics}. 

As a first step, we perform sensitivity tests by fitting the expected DM profile to background-only simulations, keeping $J_{\rm obs}$ fixed at the central value given in Table~\ref{tab:dsphs} and increasing $\sigmaobs$ in quadrature by 0.3, 0.5 and 1.0. 
In Fig.~\ref{fig:UMi-null-hyp}, we show how relaxing our assumed knowledge of the $J$-factor weakens our sensitivity to $\sigmav$ from Ursa Minor. If we include only the uncertainties in Ursa Minor’s $J$-factor quoted in Table~\ref{tab:dsphs} ($\sigmaobs = \sigma_{\rm cat}$), the median simulation yields a 95\% C.L. upper limit of $\sigmav = 7.9\times 10^{-26}\,{\rm cm}^3/{\rm s}$. If we increase $\sigmaobs$ by 0.3, 0.5, or 1.0, the median upper limit weakens by a factor of 1.5, 2.5, or 11, respectively. Increasing $\sigmaobs$ also broadens the distribution of upper limits, consistent with the flatter shape of $\mathcal{L}_J$.

More importantly, underestimating the uncertainty associated with a given $J$-factor can artificially strengthen the resulting upper limits, potentially excluding parameter space that contains a true signal. To demonstrate this, we produced 1000 simulations of Fermi-LAT data containing a signal from a DM particle with $m_X =42\, \rm GeV$ and $\sigmav_{\rm true} = 1.4\times10^{-26}\,{\rm cm}^3/{\rm s}$, annihilating to $b\bar{b}$ (chosen to reproduce the observed characteristics of the GCE~\cite{DiMauro:2021qcf}). In each simulation, the normalization of the injected signal is controlled by $J_{\rm true}$, which is fixed to the central value reported in Table~\ref{tab:dsphs}. To simulate the uncertainty on the $J$-factor inference, we draw $\log_{10} J_{\rm obs}$ from a normal distribution centered on $\log_{10} J_{\rm true}$ and with a logarithmic width $\sigmatrue$. 

If the full uncertainty associated with a given $J$-factor determination is accurately reflected in the quoted uncertainty ($\sigmatrue=\sigmacat$), the true value of the annihilation cross section should be excluded at the 95\% confidence level in no more than 5\% of the simulations.
Our baseline result, which corresponds to the blue histogram in the left panel of Fig.~\ref{fig:UMi-true-signal}, rejects less than 0.01\% of the simulations (in this case, we performed 10,000 simulations, none of which rejected the injected signal). This overcoverage arises because the profiled likelihood does not follow Wilks' theorem due to the degeneracy between $J$ and $\sigmav$. While nominal coverage does not hold, there is a clear trend: as we increase the magnitude of the unmodeled uncertainties, the distribution of the upper limits broadens, and the probability of obtaining a constraint that excludes the true cross section increases, as illustrated by the green, red, and yellow histograms. In particular, when we increase $\sigmatrue$ in quadrature by $0.3,\, 0.5$ and $1.0$, the true signal is excluded at 95\% confidence in $1.7\%,\,8.0\%$ and $22\%$ of our simulations, respectively.

In the right panels of Fig.~\ref{fig:UMi-true-signal}, we show how this artificial strengthening of the upper limits can be mitigated by widening $\mathcal{L}_J$, such that $\mathcal{L}_J(J|J_{\rm obs},\sigmaobs)$ has $\sigmaobs = \sigmatrue$.  In other words, by accounting for the true uncertainty associated with the kinematic analysis, we recover conservative coverage in the injection test. In this treatment, the true value of $\sigmav$ is excluded in fewer than 5\% of our simulations, even as we increase the value of $\sigma_{\rm true}$.

\begin{figure}
    \centering
    \includegraphics[width=1\linewidth]{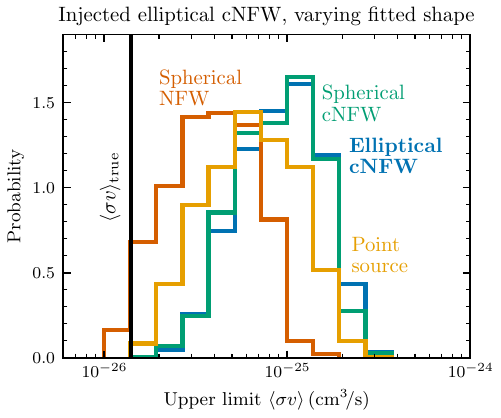}
    \caption{\textbf{GCE-like signal from Ursa Minor with variations in the spatial profile.} As in Fig.~\ref{fig:UMi-true-signal}, but for an injected DM signal generated from an elliptical cNFW profile with ellipticity of $e\sim 0.56$ and with the intermediate axis aligned with the line of sight. We fit these simulated data with the same elliptical cNFW profile (blue), a spherical cNFW profile (green), a spherical NFW profile (red), or a point source (yellow). The injected DM signal has an annihilation cross section of $\sigmav_{\rm true}=1.4\times10^{-26}\,{\rm cm}^3/{\rm s}$, and we adopt the catalog uncertainty, such that $\sigmaobs=\sigmatrue=\sigmacat$. \textit{Fitting a point-like or cusped spatial profile to a dwarf galaxy with a cored DM profile can lead to artificially stringent upper limits on the annihilation cross section.}}
    \label{fig:UMi-true-signal-shape}
\end{figure}

\begin{figure}[t]
    \centering
    \includegraphics[width=1\linewidth]{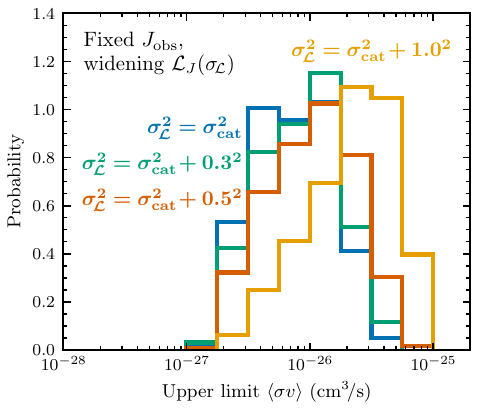}
    \caption{\textbf{Stacked sensitivity.} As in Fig.~\ref{fig:UMi-null-hyp}, but for a stacked analysis of the 27 dwarf galaxies listed in Table~\ref{tab:dsphs}. \textit{As we increase the adopted $J$-factor uncertainty to account for possible unmodeled systematics ($\sigmaobs > \sigma_{\rm cat}$), the upper limits on $\sigmav$ become weaker.}}   
    \label{fig:stacked-null-hyp}
\end{figure}

Next, we investigate the impact of the spatial morphology of the halo on DM searches in Ursa Minor. We do this by simulating datasets that include a DM signal generated from an elliptical cNFW profile with an ellipticity of $e\sim 0.56$. For simplicity, we choose the orientation such that the intermediate axis is aligned with the line of sight. The observed ellipticity is obtained by adopting $a = 1.315,\, b = 1.315$, and $c = 0.578$, corresponding to an oblate profile. In Fig.~\ref{fig:UMi-true-signal-shape}, we show the results obtained by fitting these simulated data with the same elliptical cNFW profile (blue), a spherical cNFW profile (green), a spherical NFW profile (red), or a point source (yellow). For simplicity, the three extended templates are assigned the same $J$-factor normalization and shape parameters, even though kinematic analyses of the corresponding three-dimensional profiles would generally yield different values. When fitting the emission as a point source, we modify $\mathcal{L}_J$ by replacing $J_{\rm tot}$ with $J(0.5^\circ)$, following Ref.~\cite{DiMauro:2022hue}. 

The results shown in Fig.~\ref{fig:UMi-true-signal-shape} demonstrate that ellipticity does not have a strong impact on the resulting fit, consistent with the results of Ref.~\cite{Klop:2016lug}. The concentration and spatial extent of the halo profile, however, can have a significant impact. The significance of the signal decreases when a spherical NFW profile is adopted, while the median upper limit becomes stronger by a factor of $2.4$. By contrast, fitting the dwarf as a point source yields a slightly higher TS while strengthening the median limit by a factor of $1.4$, consistent with Ref.~\cite{DiMauro:2022hue}. In other words, treating extended dwarf galaxies as point-like sources can lead to artificially stringent constraints in standard Fermi-LAT analyses.

\begin{figure*}
    \centering
    \includegraphics[width=1\linewidth]{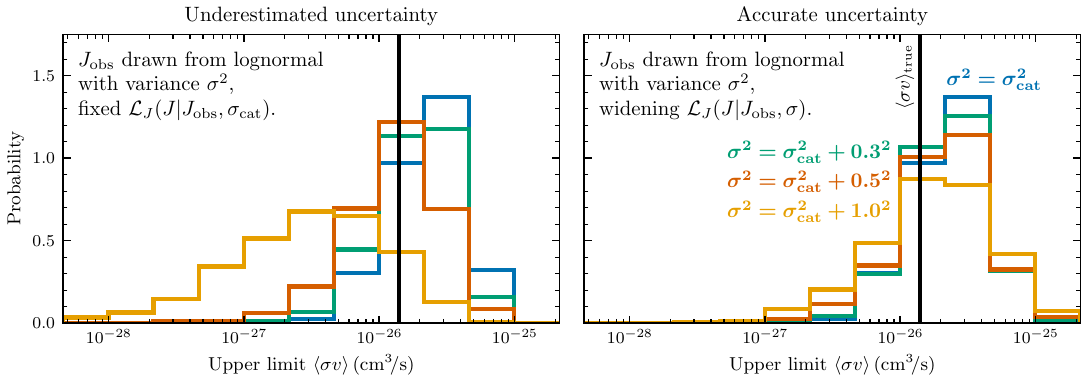}
    \caption{\textbf{GCE-like signal from a stacked sample of dwarf galaxies with increased uncertainty.} 
    As in Fig.~\ref{fig:UMi-true-signal}, but for a stacked analysis of the 27 dwarf galaxies listed in Table~\ref{tab:dsphs}. \textit{In a stacked analysis, the impact of an unmodeled systematic uncertainty common to all $J$-factor determinations is more severe than in an analysis of a single dwarf galaxy and is more difficult to mitigate.}}
    \label{fig:stacked-true-signal}
\end{figure*}

\section{The Impact of Unmodeled Uncertainties on Stacked Analyses of Dwarf Galaxies}
Indirect searches for DM annihilation in dwarf galaxies achieve greater sensitivity by stacking observations of many such systems – that is, by replacing the individual likelihoods in Eq.~\ref{eq:Lone} with the joint likelihood in Eq.~\ref{eq:Ljoint}. In this Section, we jointly analyze the simulations of the 27 dwarf galaxies listed in Table~\ref{tab:dsphs}.

Fig.~\ref{fig:stacked-null-hyp} shows that, for fixed values of $J_{\rm obs}$ and $\sigmaobs=\sigmacat$, stacking the likelihoods improves the sensitivity by an order of magnitude relative to the analysis of Ursa Minor alone, yielding a median upper limit of $\sigmav= 7.5\times 10^{-27}\, {\rm cm}^3/{\rm s}$. This improvement results from the statistical power gained by combining many different dwarf galaxies, including several with large $J$-factors. Dwarf galaxies with the largest $J$-factors typically yield the highest signal-to-noise ratios and have the greatest influence on the resulting constraints.

As shown in the previous section for a single dwarf galaxy, Fig.~\ref{fig:stacked-null-hyp} demonstrates that increasing the unmodeled uncertainties in the $J$-factors ($\sigmaobs > \sigmacat$) weakens the upper limits on $\sigmav$ and broadens their distribution. In particular, increasing $\sigmaobs$ in quadrature by 0.3, 0.5, and 1.0 weakens the median upper limit by a factor of 1.14, 1.44, and 3.23, respectively. This effect is considerably weaker than in the analysis of Ursa Minor alone. The primary reason is that, as shown in Fig.~\ref{fig:Jcomparison}, the catalog uncertainty for Ursa Minor is much smaller than those of the dwarf galaxies with the largest $J$-factors. Consequently, adding the same uncertainty in quadrature produces a smaller fractional increase in $\sigmaobs$ for the stacked analysis.

In Fig.~\ref{fig:stacked-true-signal}, we show the distribution of upper limits obtained from the stacked analysis after injecting a DM annihilation signal with $\sigmav=1.4\times 10^{-26}\, {\rm cm}^3/{\rm s}$. As in the previous Section, $J_{\rm true}$ is used to normalize the injected signal. The likelihood has width $\sigma_{\mathcal L}$ and is centered on $J_{\rm obs}$, which is independently drawn from a log-normal distribution centered on $J_{\rm true}$ with logarithmic width $\sigma_{\rm true}$. When the uncertainties quoted in Table~\ref{tab:dsphs} are adopted for both $\sigmaobs$ and $\sigmatrue$, we find that the true signal is excluded at the 95\% confidence level in approximately 21\% of the realizations. This significant undercoverage arises because the number of nuisance parameters in our fit – namely, the $J$-factors -- grows linearly with the number of dwarf galaxies. The asymptotic result provided by Wilks' theorem, which we use to define the 95\% upper limit through $\Delta{\rm TS}=2.71$, assumes that the number of nuisance parameters remains fixed and small relative to the sample size. These conditions are not satisfied in our stacked analysis of dwarf galaxies. Ref.~\cite{McDaniel:2023bju}, for example, addresses this undercoverage by using blank fields to calibrate the analysis.

Although these limits do not have the nominal coverage, the left panel of Fig.~\ref{fig:stacked-true-signal} demonstrates that underestimating the uncertainties in the $J$-factors can make the apparent exclusion of signals present in the data even more severe. In particular, when we fix $\sigmaobs=\sigmacat$ and increase $\sigmatrue$ in quadrature by $0.3,\,0.5$, and $1.0$, the 95\% upper limits exclude the true signal in 31\%, 49\%, and 88\% of the realizations, respectively. This behavior is consistent with, but considerably more pronounced than, that found in the analysis of Ursa Minor alone. In the right panel of Fig.~\ref{fig:stacked-true-signal}, we show that increasing $\sigmaobs$ to match $\sigmatrue$ does not fully compensate for the increased dispersion in $J_{\rm obs}$. The true signal is then excluded at the 95\% confidence level in 23\%, 28\%, and 38\% of the realizations, respectively. 

There are two reasons for this undercoverage. First, the values of $J_{\rm obs}$ are sampled from log-normal distributions. Although these distributions are symmetric in $\log_{10} J_{\rm obs}$, they have long upper tails in $J_{\rm obs}$. Upward fluctuations can dominate the resulting constraints, while downward fluctuations generally do not carry enough statistical weight to compensate for them. Second, in a stacked analysis of 27 independent dwarf galaxies, the probability that at least one has an upward fluctuation greater than $1.5\sigma$ is approximately $0.85$. Upward outliers with large values of $J_{\rm obs}$ can therefore produce artificially stringent constraints on $\sigmav$. Increasing $\sigma_{\mathcal L}$ broadens the $J$-factor likelihoods and relaxes these constraints, but it does not eliminate the impact of upward fluctuations in $J_{\rm obs}$.

As a final note, upward fluctuations in the case of $\sigmatrue^2=\sigmacat^2+1.0^2$ can produce extremely large $J$-factors, $\log_{10} J \gtrsim 21$, which are expected to be exceedingly unlikely according to cosmological simulations~\cite{Raman:2026zbr}. Incorporating cosmological priors into the analysis could therefore provide a safeguard against such extreme values. In any case, the left panel of Fig.~\ref{fig:stacked-true-signal} demonstrates the importance of accurately estimating the uncertainties associated with $J$-factor determinations.

\begin{figure}
    \centering
    \includegraphics[width=1\linewidth]{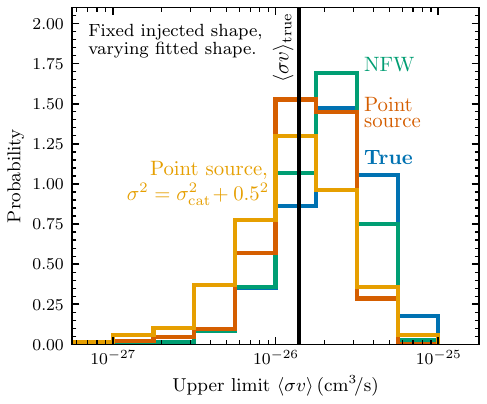}
    \caption{\textbf{GCE-like signal from a stacked analysis with variations in the spatial profiles.} As in Fig.~\ref{fig:UMi-true-signal}, but for a stacked analysis in which the injected DM signal is generated using cNFW profiles for the classical dwarfs and NFW profiles for the ultra-faint dwarfs. The injected signal has an annihilation cross section of $\sigmav_{\rm true} = 1.4\times 10^{-26}\, {\rm cm}^3/{\rm s}$. Except where otherwise specified, we adopt the catalog uncertainties, such that $\sigmatrue = \sigmaobs =\sigmacat$. We fit the simulated data using the true profile for each dwarf (blue), NFW profiles for all dwarfs (green), or point-source models for all dwarfs (red). The yellow histogram shows the case in which all dwarfs are treated as point sources, while the true $J$-factor variance is increased to $\sigmatrue^2 =\sigmacat^2+0.5^2$ and the uncertainty adopted in the likelihood remains fixed at $\sigmaobs = \sigmacat$. \textit{Mismodeling the halo profiles and $J$-factors of dwarf galaxies can substantially increase the probability that a stacked analysis will artificially exclude a signal that is present in the data.}}
    \label{fig:stacked-true-signal-shape}
\end{figure}

Finally, in Fig.~\ref{fig:stacked-true-signal-shape}, we show how mismodeling the spatial profiles of the DM halos can artificially strengthen the upper limits on $\sigmav$. In this test, we inject a true DM signal with $\sigmav = 1.4\times 10^{-26}\, {\rm cm}^3/{\rm s}$. For each dwarf, we draw $J_{\rm obs}$ from a log-normal distribution centered on $J_{\rm tot}$ listed in Table~\ref{tab:dsphs}, with $\sigmatrue = \sigmacat$, and use this value to normalize the injected signal. For the classical dwarf galaxies, we simulate the data using cNFW profiles, whereas for the ultra-faint dwarfs, we use NFW profiles. We then calculate the limits obtained by fitting these data to different choices of the halo profile. 

The blue histogram shows the results obtained when every dwarf galaxy is fit using its true halo profile. As discussed above, the true signal is excluded in 21\% of the realizations. For the green histogram, the classical dwarfs are instead fit with NFW profiles, increasing the exclusion rate to 24\%. For the red histogram, all dwarf galaxies are treated as point sources, and the true signal is excluded in 37\% of the realizations. Finally, the yellow histogram shows the limits obtained when all dwarfs are treated as point sources and the true $J$-factor uncertainty is increased to $\sigmatrue^2 = \sigmacat^2+0.5^2$, while the uncertainty adopted in the likelihood remains fixed at $\sigmaobs = \sigmacat$. In this case, the true signal is excluded in 49\% of the realizations. Because the point source fits use the smaller $J(0.5^\circ)$ rather than $J_{\rm tot}$, extreme upward fluctuations to large values of $J_{\rm obs}$ occur less frequently, reducing the impact of increasing $\sigmatrue$.

\section{Summary and Conclusions}

In this work, we have reviewed the principal sources of systematic uncertainty in determinations of the $J$-factors of Milky Way dwarf galaxies. These include uncertainties associated with stellar membership and unresolved binary populations, assumptions regarding the stellar and DM density profiles, departures from spherical symmetry and dynamical equilibrium, and the neglect of spatial extension in gamma-ray analyses. As illustrated in Fig.~\ref{fig:Jcomparison}, the dispersion among $J$-factor determinations reported in the literature is often significantly larger than would be expected from their quoted uncertainties. Taken together, these results suggest that an additional uncertainty of order $\Delta\log_{10}J\sim0.5$ may be a conservative and appropriate choice for many dwarf galaxies.

We used simulated Fermi-LAT observations to quantify the impact of these uncertainties on searches for DM annihilation. We parameterized the uncertainties in the $J$-factor determinations using two quantities: $\sigma_{\mathcal L}$, which controls the width of the $J$-factor likelihood adopted in the gamma-ray analysis, and $\sigmatrue$, which controls the actual dispersion of the inferred $J_{\rm obs}$ around the true central value. This distinction allowed us to study both the effect of increasing the assumed uncertainty and the consequences of adopting an uncertainty that is smaller than the actual uncertainty in the kinematic inference of the $J$-factor.

For the case of a single dwarf galaxy (Ursa Minor), we found that increasing $\sigmaobs$ in quadrature by 0.3, 0.5, and 1.0 weakens the median upper limit on $\sigmav$ by factors of 1.5, 2.5, and 11, respectively. More importantly, when the true uncertainty is underestimated ($\sigmatrue>\sigmaobs$), an injected signal can be excluded substantially more often than implied by the nominal confidence level. This bias can be mitigated by increasing $\sigmaobs$ to match the true underlying uncertainties, $\sigmatrue$, restoring conservative coverage in the analysis of Ursa Minor.

The consequences are more severe in the stacked analysis of 27 dwarf galaxies. Even in the baseline case, with $\sigmaobs=\sigmacat$, the nominal 95\% confidence-level upper limit can exclude signals present in the data in far more than 5\% of realizations. Thus, the standard asymptotic likelihood construction does not provide the expected coverage for this analysis. This motivates calibrating stacked analyses using simulations or blank fields, as done in Refs.~\cite{McDaniel:2023bju,Circiello:2026inp}, rather than relying exclusively on asymptotic likelihood thresholds.  When the $J$-factor uncertainties are also underestimated ($\sigmatrue>\sigmaobs$), the undercoverage becomes considerably more pronounced. Although widening the $J$-factor likelihoods reduces this bias, setting $\sigmaobs=\sigmatrue$ does not fully restore the nominal coverage in the stacked analysis. This residual effect arises in part because upward fluctuations in the log-normally distributed $J$-factors can disproportionately influence the joint constraint. 

We also studied the consequences of mismodeling the spatial distributions of the dwarf galaxy halos. For Ursa Minor, fitting an injected cNFW signal with a spherical NFW profile artificially strengthens the median upper limit by a factor of 2.4, while treating the dwarf as a point source strengthens it by a factor of 1.4. In the stacked analysis, treating all dwarf galaxies as point sources significantly increases the rejection rate of injected signals. These results are consistent with Ref.~\cite{DiMauro:2022hue} and motivate incorporating spatial extension and halo morphology as nuisance parameters in searches for DM annihilation from dwarf galaxies.

These findings are particularly relevant to the interpretation of the GCE. Current dwarf galaxy searches are beginning to probe the range of annihilation cross sections favored by a DM interpretation of the excess. Our results show, however, that the apparent tension between the Galactic Center signal and dwarf galaxy upper limits can be overstated if the $J$-factor uncertainties, spatial extension, or halo profiles are not modeled accurately, or if the quoted limits do not have the nominal statistical coverage. This does not diminish the importance of dwarf galaxies as targets. Rather, it emphasizes that their ability to confirm or exclude a DM interpretation of the excess relies on a reliable treatment of their astrophysical and statistical uncertainties.

The Fermi Gamma-Ray Space Telescope will continue to collect data and, in the more distant future, telescopes such as the proposed Advanced Particle-astrophysics Telescope (APT)~\cite{Alnussirat:2021tlo,APT:2021lhj} will be far more sensitive to DM annihilation in dwarf galaxies~\cite{Xu:2023zyz}. In parallel to these gamma-ray observations, the Rubin Observatory and other surveys are expected to discover additional dwarf galaxies. These developments should improve the sensitivity of gamma-ray dwarf galaxy analyses~\cite{He:2013jza,LSSTDarkMatterGroup:2019mwo,Fermi-LAT:2016afa,Hargis:2014kaa}. Cosmological simulations suggest, however, that the number of undiscovered dwarfs with exceptionally large $J$-factors may be limited~\cite{Raman:2026zbr}. Future constraints may therefore remain dominated by systems that are already known, making improved stellar spectroscopy, more flexible dynamical modeling, realistic spatial templates, physically motivated priors, and direct coverage calibration increasingly important. Regardless of whether future sensitivity gains arise from deeper gamma-ray exposure or newly discovered targets, robust constraints on DM annihilation will require the uncertainties associated with dwarf galaxy $J$-factor determinations to be brought under substantially better control.

\bigskip
\bigskip

\textit{Acknowledgments}--- We would like to thank Josh Foster for helpful discussions. This work has been supported by the Office of the Vice Chancellor for Research at the University of Wisconsin-Madison, with funding from the Wisconsin Alumni Research Foundation. AKZ is further supported
by the National Science Foundation under grants PHY-2209445 and OPP-2042807 and by the Balzan Foundation. We disclose the use of Gemini 3.6 for code development.

\bibliography{biblio}

\end{document}